# Modeling the Effects of Velocity, Spin, Frictional Coefficient, and Impact Angle on Deflection Angle in Near-elastic Collisions of Phenolic Resin Spheres

S.C. Crown[1]

*Sam's Applied Physics & Trucking Co.*

*Gresham, OR 97080, USA*

______________________________________________________

## Abstract

A simple model is outlined to describe the collision of cast phenol-formaldehyde resin spheres such as the balls used in the parlor game of pocket billiards, based in part on the famous analysis of elastic collisions developed by Heinrich Hertz over 100 years ago. The analysis treats the normal and tangential components of the initial sphere's velocity independently as it collides with a stationary identical second sphere. The collective effects of these and other parameters on the trajectory of the second sphere are provided in the conclusions.

## 1. Introduction

This work was motivated primarily by an interest in developing an improved understanding of the physics involved in the collision of billiard balls to supplement an intuitive, experience-based understanding of these phenomena and to while away several rainy Oregon winter afternoons. For quasi-elastic contact (the coefficient of restitution approaches unity), force-displacement mechanics provide a reasonably approximate linear elastic model sufficient to demonstrate the physics of the process.

In most collisions, plastic deformation and viscous friction dissipate energy that results in coefficients of restitution less than unity, and the collision / contact problem then becomes so complex that an accurate theoretical solution is difficult to obtain. In the present analysis, with physical conditions as stated below, a near approximation of the actual collision may be achieved by treating the collision as perfectly elastic. The collision event is modeled between two elastically deformable spheres ($S_1$ and $S_2$) colliding with normal and tangential momentum components. In the latter case, friction plays a prominent role in the coefficient of restitution; see Stronge [5] and Vu-Quoc *et al*. [6] for further details.

a. *Definitions and initial conditions*

For the purposes of this paper, the following initial conditions shall apply:

1) The collision event begins at time $t = 0$ at first contact, and ends when the spheres separate at time $t = \tau$.

2) The centers of the two touching spheres at time $t = 0$ shall define the x-axis.

3) The point of contact between $S_1$ and $S_2$ at time $t = 0$ shall define the y-axis.

---

[1] Driver, loader, & chief scientist; Sam's Applied Physics & Trucking Co. *E-mail address:* samcrown @ earthlink.net

4) The initial velocity of $S_1$ is defined as $v$ and is directed at an obliquity angle $\alpha$ above the x-axis. The normal and tangential components of $v$ are respectively

$$v_x = v \cos(\alpha) \tag{1.1}$$

and

$$v_y = v \sin(\alpha) \,. \tag{1.2}$$

In addition, the pre-impact spin of $S_1$ is $\beta_{10}$.

5) Movement of the two spheres is constrained to a frictionless horizontal plane.

b. The following physical parameters are defined for the purposes of the example to be developed:

1) Radius of $S_1$ and $S_2$:
$$r = 2.857 \times 10^{-2} \,\mathrm{m} \tag{1.3}$$

2) Mass of each sphere:
$$M = 1.701 \times 10^{-1} \,\mathrm{kg} \tag{1.4}$$

3) Modulus of elasticity:
$$E = 5.84 \times 10^{9} \,\mathrm{Pa} \tag{1.5}$$

4) Coefficient of friction (clean):
$$\mu = 8 \times 10^{-1} \tag{1.6}$$

5) Poisson's ratio
$$\gamma = 3.4 \times 10^{-1} \tag{1.7}$$

## 2. The normal component

During a normal elastic collision, conservation of kinetic energy and momentum demand that the impulse $F \cdot \tau/2$ acting between the two identical spheres causes $S_1$ to stop relative to $S_2$ at time $t = \tau/2$.[2] When system momentum and energy are conserved and the spheres' masses are equal, and excluding the effects of the y-axis velocity component and pre-impact spin (see Section 3), both spheres continue at $t = \tau/2$ in the positive x direction with velocity $v_x/2$.

At this point in time, the structure of both spheres is elastically deformed as shown diagrammatically at Fig.1 below. Momentum and energy pertaining to the two spheres and the system center of mass during contact are tabulated below in Table 1. During the time interval $\tau/2 < t < \tau$, the deformed materials at the impact site relax, exerting a force F (see Equations 2.19 and 2.20) between the two spheres. In an elastic collision, this impulse, $F \cdot \tau/2$, is exactly equal to the impulse acting on the spheres during the interval $0 < t < \tau/2$, and its action on $S_1$, which has velocity $v_x/2$ at time $\tau/2$, further reduces $S_1$'s velocity to zero, and accelerates $S_2$'s velocity to $v_x$ at time $t = \tau$.

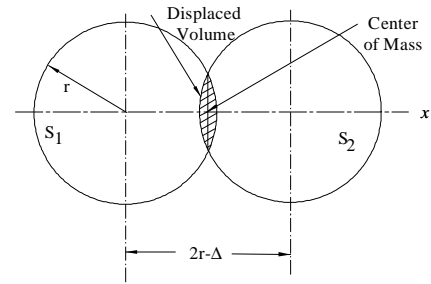

Fig. 1. The collision at time $t = \tau/2$. The shaded volume is elastically displaced by the energy of the collision. Relative velocity between $S_1$ and $S_2$ is zero, and the velocity of the center of mass is $v_x/2$.

|            | $t = 0$    | $t = \tau/2$ | $t = \tau$ |
|------------|------------|--------------|------------|
| $P_{S1}$   | $Mv_x$     | $Mv_x/2$     | 0          |
| $P_{S2}$   | 0          | $Mv_x/2$     | $Mv_x$     |
| $P_{CM}$   | $2Mv_x/2$  | $2Mv_x/2$    | $2Mv_x/2$  |
| $KE_{S1}$  | $Mv_x^2/2$ | $Mv_x^2/8$   | 0          |
| $KE_{S2}$  | 0          | $Mv_x^2/8$   | $Mv_x^2/2$ |
| $\xi$      | 0          | $Mv_x^2/4$   | 0          |

Table 1. Summary of momentum and energy[3] at three points in time during a normal collision event.

By inspection, given that the elastic deformation of the two spheres is at its maximum at time $t = \tau/2$; and that the impact plane is coincident with the center of mass of the two spheres;

---

[2] Assuming that the material is isotropic and the deformation time is equal to the relaxation time.

[3] $\xi$ is potential energy stored in the displaced strained material of $S_1$ and $S_2$.

and that the law of conservation of momentum applies to the center of mass, which has a constant velocity $v/2$; the magnitude of the maximum deflection, $\Delta$, is[4]

$$\Delta := \frac{v_x \cdot \tau}{2} \tag{2.1}$$

Applying the theoretical principles developed by Hertz[5] [1], and modified by Landau[6] and Lifshitz [2], the period that the two spheres are in contact, $\tau$, is expressed as:

$$\tau := 2.94 \left( \frac{M'^2}{k^2 \cdot v_x} \right)^{\frac{1}{5}} \tag{2.2}$$

where M' is the so-called reduced mass

$$M' := \frac{M_1 \cdot M_2}{M_1 + M_2} \tag{2.3}$$

and

$$k := 8 \cdot \frac{E}{15 \left(1 - \gamma^2\right)} \sqrt{\frac{r}{2}} \tag{2.4}$$

The force acting between the two spheres may be calculated as the product of the area $\Omega$ of the contact surface and the effective elastic modulus; and the contact area $\Omega$ is approximately a linear function of $\delta$ (see Fig. 2).

$$\Omega = \pi \cdot a^2 \tag{2.5}$$

$$a^2 = r^2 - (r - \delta)^2 \tag{2.6}$$

Expanding Eq. 2.6,

$$a^2 = r \cdot \delta \cdot \left(1 - \frac{\delta}{2 \cdot r}\right) \tag{2.7}$$

For the purposes of this discussion, Eq. (2.7) may be approximated[7] as

$$a^2 = r \cdot \delta \tag{2.8}$$

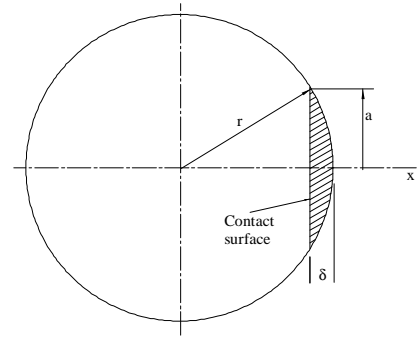

Fig.2. Geometry of one sphere's contact surface when $t = t/2$

Recognizing that the force acting between the two spheres begins at $t = 0$ and ends at $t = \tau$ and reaches its maximum amplitude of at $t = \tau/2$, and noting that the force is a linear function of $\delta$ (see Fig. 2),

$$F = \kappa \cdot \delta \tag{2.9}$$

After applying Newton's equation, Eq. 2.9 results in the differential equation for the dynamics of the impact surface relative to the sphere's center

$$\frac{d^2}{dt^2} \delta = A \cdot \sin(\omega \cdot t) \tag{2.10}$$

where A is a constant to be determined and, by inspection,

$$\omega := \frac{\pi}{\tau} \tag{2.11}$$

Integrating Eq. 2.10 and applying boundry values from Table 1,

---

[4] The symbol := designates "equal by definition"; the subscript in "$v_1$" pertains to sphere #1. The "prime" device will be used to designate a value at the conclusion of the collision, *e.g.*, "$v'_1$" identifies the post-impact velocity of sphere #1.

[5] Hertz's original analysis addressed the collision of a sphere and a perfectly rigid planar surface.

[6] Landau and Lifshitz's work accommodated the geometry attending to the collision of two spheres.

[7] The approximation introduces an error $\varepsilon < 0.5\%$

$$v_{x1} := \frac{v_x}{2} \cdot (1 + \cos(\omega \cdot t))$$

(2.12)

$$x_1 := -r + \frac{v_x}{2} \cdot \left( t + \frac{\sin(\omega \cdot t)}{\omega} \right)$$

(2.13)

Similarly,

$$v_{x2} := \frac{v_x}{2} \cdot (1 - \cos(\omega \cdot t))$$

(2.14)

$$x_2 := r + \frac{v_x}{2} \cdot \left( t - \frac{\sin(\omega \cdot t)}{\omega} \right)$$

(2.15)

Taking the first derivative of the velocities gives the time dependent expressions for the spheres' accelerations A:

$$A_{x1} := \frac{-\omega}{2} \cdot v_x \sin(\omega \cdot t)$$

(2.16)

$$A_{x2} := \frac{\omega}{2} \cdot v_x \sin(\omega \cdot t)$$

(2.17)

and the time dependent expressions for the normal forces acting on the spheres are:

$$F_{x1} := \frac{-M \cdot \omega \cdot v_x}{2} \cdot \sin(\omega \cdot t)$$

(2.18)

$$F_{x2} := \frac{M \cdot \omega \cdot v_x}{2} \cdot \sin(\omega \cdot t)$$

(2.19)

As expected, equal and opposite forces are acting on $S_1$ and $S_2$. The force $F_1$ acts on $S_1$ in the negative x direction, and the force $F_2$ is acting on $S_2$ in the positive x direction.

## 3. The tangential component

As noted previously, collisions that occur when $S_1$'s velocity vector is not congruent with a straight line connecting the centers of the two spheres are characterized as "oblique" and are distinguished from normal impacts by the addition of a tangential velocity component of magnitude $v \cdot \sin(\alpha)$ plus $S_1$'s surface velocity due to its angular velocity (see Fig.4). The angle between v and the x-axis is characterized as the obliquity angle $\alpha$ (see Fig. 3). Unlike the case addressing the normal component discussed in the previous section where the forces and quasi-elastic deformation of the spheres during the collision are axisymmetric, during a collision where a tangential velocity component is present, the tangential momentum of $S_1$ introduces a corresponding equal and opposite non-symmetrical shear strain into the material of both spheres at the impact site (see Fig. 5). At the energies pertinent to this analysis, this strain is less than the elastic limit of the material as demonstrated by the fact that pool balls do not sustain permanent deformation during normal play. Nevertheless, its presence introduces complexity into the collision process that is poorly understood and is currently the subject of significant research. Also, the presence of shear strain obscures the volume of displaced material due only to the normal component, and the presence of angular momentum in oblique collisions adds further complexity to their description. With the knowledge that doing so will introduce error into the following discussion, I shall nevertheless entertain the untested assumption that the dynamic normal and tangential forces are independent at the energies cited here, *i.e.* each unaffected by the other, as articulated by Castigliano's first theorem[8]. Given the physical properties and energies described above, the induced error is expected to be negligible for the purposes of this paper.

Since both spheres are constrained to a horizontal, frictionless plane, they accordingly move without rolling or otherwise rotating except during the collision event. However, during play of pocket billiards, it is common practice for the player to deliberately introduce spin to $S_1$ prior to its impact with $S_2$ with the intent of influencing $S_1$'s and $S_2$'s post impact trajectories. For the purposes of this discussion, this spin vector is assumed to be normal to the horizontal playing surface, which is not always the case in practice.

---

[8] Castigliano's first theorem states that deflection is the partial derivative of strain energy with respect to any one of the applied forces *in a statically loaded structure* and is equal to the displacement of the point of application of that force. See [8], p.51

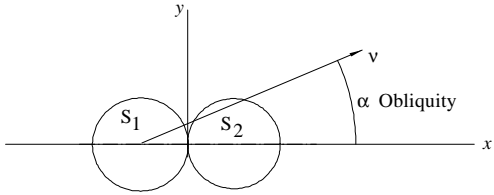

Fig. 3. Obliquity is the angle $\alpha$ between $S_1$'s initial velocity and the x axis connecting the centers of the spheres at impact..

During the period of contact, $0 \leq t \leq \tau$, the forces acting at the impact site are the normal forces (Eqs. 2.18 and 2.19), and tangential forces[9], see Fig. 4,

$$F_y = \mu \cdot F_n = M \cdot \frac{d}{dt} v_y = \frac{I}{r} \cdot \frac{d^2}{dt^2} \theta$$

(3.1)

where I is the spheres' moment of inertia, and $\mu F_n$ is the retarding frictional force. Fig. 5 graphically describes the effects of the tangential velocity components during the collision. At first contact, each spheres is tangentially accelerated as shown.

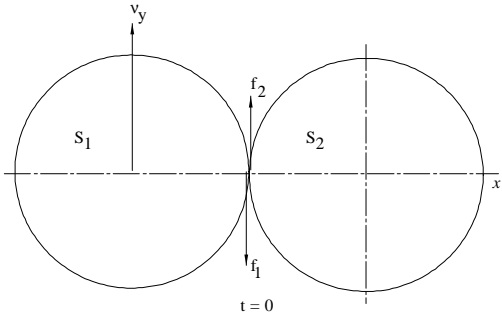

Fig. 4. During the collision, equal and opposite tangential forces[10] may be applied to $S_1$ and $S_2$.

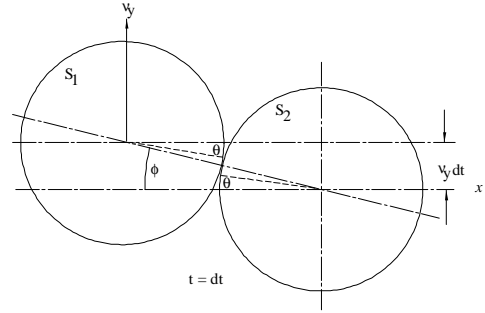

Fig. 5. At time t = dt, each sphere has rotated clockwise through angle $d\theta$. The angular displacement of $S_1$ relative to $S_2$ is designated as $\phi$. Note that $r\phi > r\theta$ when $\beta = 0$.

The coefficient of dynamic friction[11] between the two uncontaminated surfaces is $\mu$, and the moment of inertia of each sphere about its center of mass is

$$I := \frac{2}{5} \cdot M \cdot r^2$$

(3.2)

The tangential forces acting in opposite directions on $S_1$ and $S_2$ at the point of contact are

$$M \cdot \frac{d^2}{dt^2} y_1 = -\ni \cdot \mu \cdot F_{x1} = f_1$$

(3.3)

and

$$M \cdot \frac{d^2}{dt^2} y_2 = \ni \cdot \mu \cdot F_{x2} = f_2$$

(3.4)

where

$$\ni = 1 \quad \text{when } r \cdot \beta_{10} + v_y \geq 0$$

(3.5)

$$\ni = -1 \quad \text{when } r \cdot \beta_{10} + v_y < 0 \;;$$

(3.6)

the initial angular velocities (in radians/sec) of the spheres[12] are designated as $\beta_{10}$ and $\beta_{20}$; and the other initial conditions at time t = 0 are: the y-axis velocity of $S_1$ is $v_y$; the y-axis velocity of $S_2$ is 0; and the position of the point of contact at time t = 0 is x = 0, y = 0. Solving equations 3.4 and 3.5 for the y-axis velocities,

$$v_{y1} := v_y - \frac{\ni \cdot \mu \cdot v_x}{2} \cdot (1 - \cos(\omega \cdot t)) \tag{3.7}$$

$$v_{y2} := \frac{\ni \cdot \mu \cdot v_x}{2} \cdot (1 - \cos(\omega \cdot t)) \tag{3.8}$$

---

[9] The direction and magnitude of $F_1$ and $F_2$ in Fig. 4 are for illustration only; under some conditions, they may be reversed, or zero.

[10] Torque vectors $\Gamma_1 = r \times f_1$, and $\Gamma_2 = r \times f_2$ are perpendicular to and directed into the plane of the diagram when $r \times \beta < v_{1y}$. These tangential forces are the effort required to overcome the frictional drag as one surface slides over the other. When the surface velocities of the two spheres are equal there is no slipping and the tangential force disappears.

[11] As a practical matter, we should recognize that the presence of cheeseburger grease on the spheres in pool joints significantly reduces $\mu$.

[12] The notation "$\beta_{10}$" simply identifies the angular velocity of $S_1$ at time t = 0.

The angular dynamic of each sphere is

$$I \cdot \frac{d^2}{dt^2} \theta = \ni \mu \cdot r \cdot F_{x1} \tag{3.9}$$

The surface velocities V of the two spheres are

$$V_1 := v_{y1} + r \cdot \beta_1 \tag{3.10}$$

where β, the angular velocity is,

$$\beta_1 := \beta_{10} - \frac{5 \cdot \ni \cdot \mu \cdot v_x \cdot (1 - \cos(\omega \cdot t))}{4 \cdot r} \tag{3.11}$$

and

$$V_1 := (v_y + r \cdot \beta_{10}) - \frac{7}{4} \cdot \ni \cdot \mu \cdot v_x \cdot (1 - \cos(\omega \cdot t)) \tag{3.12}$$

Likewise[13],

$$\beta_2 := -\frac{5 \cdot \ni \cdot \mu \cdot v_x \cdot (1 - \cos(\omega \cdot t))}{4 \cdot r} \tag{3.13}$$

and

$$V_2 := \frac{7}{4} \cdot \ni \cdot \mu \cdot v_x \cdot (1 - \cos(\omega \cdot t)) \tag{3.14}$$

The condition for slipping to cease between the two surfaces is that their surface velocities at the impact site are equal, *i.e.*,

$$V_1 - V_2 = 0 \tag{3.15}$$

or that they continue to slip throughout the period of the collision. Solving Eq. 3.15 for the point in time when the surface velocities are equal[14], and the tangential forces acting on the two spheres equal zero[15].

$$T := \frac{\tau}{\pi} \cdot \text{acos}\left[1 - \frac{2(v_y + r \cdot \beta_{10})}{7 \cdot \ni \mu \cdot v_x}\right] \tag{3.16}$$

When the complement of the inverse cosine term in Equation 3.16 has the property that

$$\frac{v_y + r \cdot \beta_{10}}{\mu \cdot v_x} > 7 \tag{3.17}$$

the solution to Equation 3.16 will have an imaginary component, which is interpreted as an indication that the two surfaces continue to slip throughout the collision. Accordingly,

$$T = \frac{\tau}{\pi} \cdot \text{acos}\left[1 - \frac{2}{7} \cdot \left(\frac{v_y + r \cdot \beta_{10}}{\mu \cdot \ni \cdot v_x}\right)\right] \tag{3.18}$$

when

---

[13] Since $\beta_{20}$ is always zero for the purposes of this paper, it is omitted for the remainder of this discussion.

[14] The notation "acos" denotes the inverse cosine function.
[15] Assuming that no residual y-axis strain energy or dynamic torsional oscillation resides in either of the two spheres at this time

$$\frac{v_y + r\beta_{10}}{\mu \cdot \ni \cdot v_x} \leq 7$$

(3.19)

and

$$T = \tau \quad \text{when} \quad \frac{v_y + r\beta_{10}}{\mu \cdot \ni \cdot v_x} > 7$$

(3.20)

The computation of the spheres' y-axis and angular terminal velocities is performed on this basis since no further y-axis forces are present after $t = T$. Accordingly,

$$v'_{y1} := v_y - \frac{\ni \cdot \mu \cdot v_x}{2} \cdot (1 - \cos(\omega \cdot T))$$

(3.21)

$$v'_{y2} := \frac{\ni \cdot \mu \cdot v_x}{2} \cdot (1 - \cos(\omega \cdot T))$$

(3.22)

and the terminal x-axis velocities are as described at Eqs. 2.13 and 2.15 at time $t = \tau$. Similarly, the terminal angular velocities of $S_1$ and $S_2$ (see Eqs. 3.11 and 3.13) are their velocities at time $t = T$, and the y-axis coordinates for the centers of the two spheres at time $t = \tau$ are:

$$y'_1 := \int_0^T \left( v_y - \frac{\mu \cdot v_x \cdot \omega \cdot t}{2} \cdot \sin(\omega \cdot t) \right) dt + v'_{y1} \cdot (\tau - T)$$

(3.23)

$$y'_2 := \int_0^T \frac{\mu \cdot v_x \cdot \omega \cdot t}{2} \cdot \sin(\omega \cdot t) \, dt + v'_{y2} \cdot (\tau - T)$$

(3.24)

At the conclusion of the collision, $t = \tau$, the total system kinetic energy $\Psi$ is

$$\Psi = \Psi_x + \Psi_y + \Psi_\theta + \Psi_f$$

(3.25)

where $\Psi_x$ are $\Psi_y$ the energies associated with the x- and y-axis translational velocity components; $\Psi_\theta$ is the rotational energy of the two spheres, and $\Psi_f$ is the frictional heat energy generated by slipping between the spheres, where

$$\Psi_x := \frac{M}{2} \cdot \left( v'^2_{x1} + v'^2_{x2} \right)$$

(3.26)

$$\Psi_y := \frac{M}{2} \cdot \left( v'^2_{y1} + v'^2_{y2} \right)$$

(3.27)

$$\Psi_\theta := \frac{I}{2} \cdot \left( \beta'^2_1 + \beta'^2_2 \right)$$

(3.28)

Since x-axis energy is conserved, *viz.*, $v_{x2}$ at $t = \tau$ is equal to $v_{x1}$ at $t = 0$, the energy required to accelerate $S_{y2}$, generate frictional heat, and change the angular velocity of both spheres is supplied by an equivalent reduction of y-axis and / or spin energy, *i.e.*, the y-axis velocity and angular velocity of $S_1$. Also, since the angular accelerations of $S_1$ and $S_2$ are equal, as are their y-axis accelerations, and observing, from the definition of the moment of inertia, that $S_2$'s rotational and y-axis energies are related by

$$\frac{\Psi_{rotational}}{\Psi_{translational}} = \frac{5}{2}$$

(3.29)

and that the frictional energy $\psi_f$ generated during the collision is expressed as

$$\Psi_f = \int_0^T \left( V_1(t) - V_2(t) \right) \cdot \mu \cdot F_x(t) \, dt$$

(3.30)

It is a simple matter to confirm that kinetic energy is conserved.

The deflection velocity of $S_1$ is the vector sum of its post collision x- and y-axis velocity components (see Eq. 2.12 and 3.7) and is defined as,

$$v'_1 := \sqrt{v'_{x1}{}^2 + v'_{y1}{}^2} \qquad (3.31)$$

and its deflection angle relative to the x-axis is defined as

$$\Theta_1 := \mathrm{asin}\left(\frac{v'_{y1}}{v'_1}\right) \qquad (3.32)$$

Similarly, the deflection velocity and angle of $S_2$ are

$$v'_2 := \sqrt{v'_{x2}{}^2 + v'_{y2}{}^2} \qquad (3.33)$$

$$\Theta_2 := \mathrm{atan}\left(\frac{v'_{y2}}{v'_{x2}}\right) \qquad (3.34)$$

### 4. Conclusions

As described in Section 3, the degree of pre-impact spin ($\beta_{10}$) and slip between the two spherical surfaces during the collision directly effect the post- collision deflection angle and velocity of $S_2$[16], but $S_1$'s post-collision non-zero deflection $\Theta_1$ is always parallel to the y-axis and is independent of its initial velocity, spin, obliquity, and the coefficient of friction $\mu$.

The following figures graphically illustrate the effect on the deflection angle $\Theta_2$ as the various parameters discussed in the previous sections are independently varied. The parametric values of $\alpha$, $\beta_{10}$, $v$, and $\mu$ are specified on each diagram.

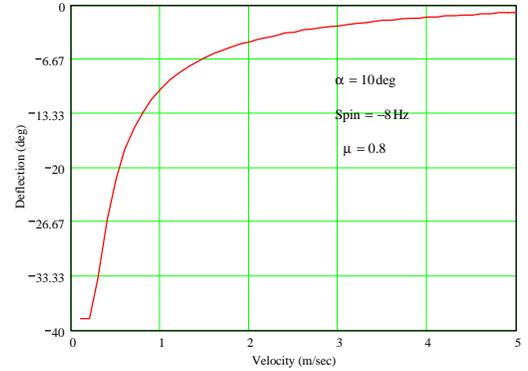

Fig. 6. Illustrating the effect of $S_1$'s initial velocity on the deflection of $S_2$. Notice that $S_2$'s deflection at low energy is greater than at higher energies. See Eq. 3.31. By inspection of Eq. 3.29, the deflection angle of $S_1$ is independent of its initial velocity.

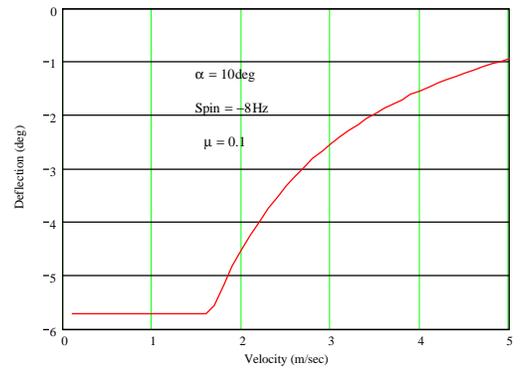

Fig. 7. Illustrating the very significant reduction in $S_2$'s deflection due to the reduced coefficient of friction. The flat portion of the curve at lower velocity is due to the reduced energy transfer from $S_1$ to $S_2$ caused by the two surfaces slipping throughout the period of the collision.

---

[16] This result contradicts the commonly held belief among casual pool players that $S_2$'s trajectory is always coincident with the x-axis. This is true when $\mu = 0$, and leads one to conclude that the conventional wisdom is influenced by the aforementioned cheeseburger grease.

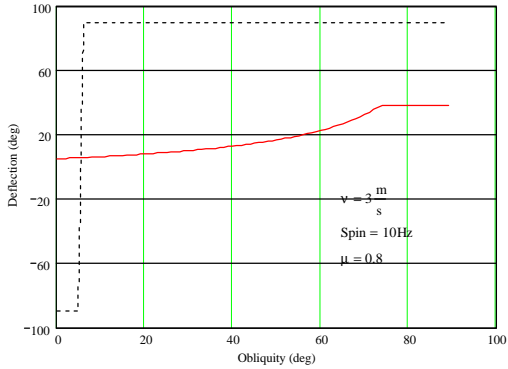

Fig. 8. Illustrating the effect of the collision's obliquity on the deflection of $S_1$ (dashed black line) and $S_2$ (red solid line). The flat section of the $S_2$ curve is due to the reduced energy transfer from $S_1$ to $S_2$ caused by the two surfaces slipping throughout the period of the collision. Note that $S_1$'s deflection changes polarity but is always parallel to the y-axis.

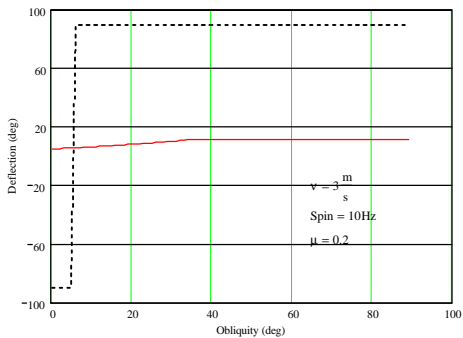

Fig. 9. Illustrating the reduction in $S_2$'s deflection due to the reduced value of $\mu$.

In general, pocket billiard players will be well advised to apply the minimum energy needed to achieve the desired results, and to remove contamination from the billiard balls when possible. This strategy will amplify the effects of spin induced to $S_1$ for the purpose of positioning it for the next shot. Alternatively, the less skilled player may opt to direct $S_1$ with higher velocity to minimize deflection error that may be induced by variability in velocity. Players should note from Figure 6 how a small variation in velocity will result in a large variation in $\Theta_2$ at lower velocities. Figure 7 suggests that the really bad player may elect to surreptitiously apply a lubricant to $S_1$ with the effect of achieving consistent deflection and confounding the more skilled opponent.

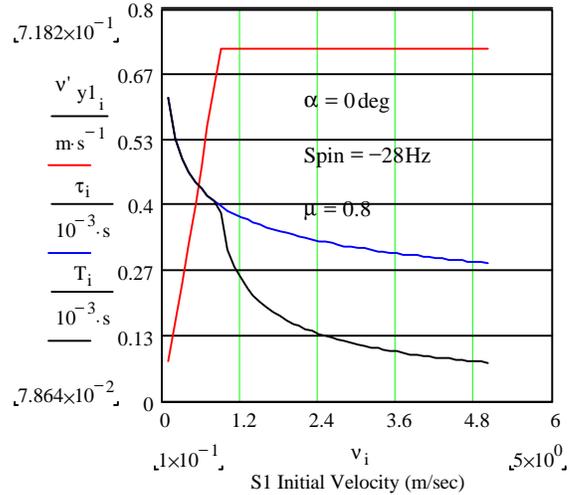

Fig. 10. Illustrating how $S_1$'s y-axis post-impact translational velocity (red) is related to the slip period T (black) and the contact period $\tau$ (blue). Note how $S_1$'s terminal velocity is limited when $T < \tau$. When $\alpha = 0$, all of $S_1$'s y-axis translational energy is derived from its pre-impact spin energy, and when the spheres cease slipping no further energy is converted.

## 5. Acknowledgments

This paper was prepared for the fun of it without assistance from anybody, except perhaps from the excellent physics faculty at Portland State University in the 1960's. The non-support of the National Science Foundation is gratefully acknowledged.

The strongest arguments prove nothing so long as the conclusions are not verified by experiment. Experimental science is the queen of sciences and the goal of all speculation.
———————————
Roger Bacon,
*Opus Tertium*